\documentclass[%
  reprint,
  groupedaddress,
  amsmath,
  amssymb,
  aps,
  prl,
  floatfix
]{revtex4-2}

\usepackage{graphicx} 
\usepackage{braket}
\usepackage{xcolor}
\usepackage{amsmath, amssymb, dsfont}
\usepackage{hyperref}
\usepackage{tikz-network} 

\tikzstyle{rounded vertex} = [draw, rectangle, rounded corners=3pt]

\definecolor{tensorblue}{RGB}{9, 132, 227}
\definecolor{tensorteal}{RGB}{0,206,201}
\definecolor{tensorpurp}{RGB}{162, 155, 254}
\definecolor{tensordpurp}{RGB}{108, 92, 231}
\definecolor{tensorpink}{RGB}{253, 121, 168}
\definecolor{tensorred}{RGB}{250,177,160}

\begin{document}

\title{
A Markovian approach to $N$-photon correlations beyond the quantum regression theorem
}
\author{Mateusz Salamon}
  \email{mateusz.salamon@manchester.ac.uk}
  \affiliation{Department of Physics and Astronomy, The University of Manchester, Oxford Road, Manchester, M13 9PL, United Kingdom}
\author{Oliver Dudgeon}
  \affiliation{Department of Physics and Astronomy, The University of Manchester, Oxford Road, Manchester, M13 9PL, United Kingdom}
\author{Ahsan Nazir}\email{ahsan.nazir@manchester.ac.uk}
  \affiliation{Department of Physics and Astronomy, The University of Manchester, Oxford Road, Manchester, M13 9PL, United Kingdom}
  \author{Jake Iles-Smith}\email{jake.iles-smith@sheffield.ac.uk}
  \affiliation{School of Mathematical and Physical Sciences, The University of Sheffield, Western Bank, Sheffield, S10 2TN, United Kingdom}
\date{\today}

\begin{abstract}
Multi-photon correlations from quantum emitters coupled to vibrational environments lie beyond the reach of standard tools such as the quantum regression theorem (QRT).
Here, we introduce a Markovian framework for computing frequency-resolved $N$-photon correlation functions that overcomes this limitation.
Applying our approach to a driven semiconductor quantum dot provides a tractable description of phonon effects on fluorescence beyond the single-photon spectrum.
Our method accurately captures the emergence of the phonon sideband, missed by conventional QRT treatments, and reveals rich phonon-induced structure in the filtered two-photon spectrum.
Strikingly, we find that photons emitted via the phonon sideband inherit second-order coherence properties of the Mollow triplet.
\end{abstract}

\maketitle
\textit{Introduction}---Correlation functions of the electromagnetic field~\cite{Glauber_1963} are fundamental tools for characterizing light in quantum optics~\cite{Carmichael_1993, Cohen-Tannoudji_1998, Vogel_2006}.
A central method for computing these functions is the Markovian quantum regression theorem (QRT)~\cite{Lax_1963, Carmichael_1993, Cohen-Tannoudji_1998, Vogel_2006, Breuer_2002}, which extends single-time evolution equations to multi-time correlation functions.
Despite its widespread use, the QRT has two key limitations: (i) it typically neglects the finite frequency resolution of physical detectors; and (ii) it is formally valid only under the standard assumption of a locally flat environmental frequency response, leading to inaccuracies or even unphysical results outside these conditions~\cite{Swain_1981, Budini_2008, Goan_2011, Guarnieri_2014, McCutcheon_2016, Cosacchi_2018, Cosacchi_2021}.

The first limitation was addressed by Eberly and Wódkiewicz with the introduction of the physical spectrum~\cite{Eberly_1977}, and later extended to frequency-resolved higher-order correlation functions~\cite{Arnoldus_1984, Knoll_1986, Cresser_1987, Vogel_2006}.
At higher orders, however, these correlation functions involve solving complex, high-dimensional integrals due to time-ordering of field operators, rendering calculations beyond second order practically intractable~\cite{Knoll_1986, Cresser_1987, Nienhuis_1993, Kamide_2015}.
This changed with the development of the sensor method by del Valle et al.~\cite{delValle_2012}, which introduces auxiliary two-level systems that act as frequency-resolved detectors of the emitted light.
The sensor method enables computation of correlation functions of arbitrary order in the absence of environmental structure, advancing significantly beyond integral-based techniques~\cite{delValle_2012, Carreno_2017, Nieves_2018}.

In contrast, the restriction of the QRT to flat environments remains a major limitation.
For example, in solid-state and molecular emitters, the vibrational environment is structured and plays a central role in determining dynamical and optical properties~\cite{Besombes_2001, Krummheuer_2002, PhysRevLett.91.127401, Machnikowski_2004, PhysRevLett.98.227403, McCutcheon_2010, Roy_2011, PhysRevLett.110.217401, Mork_2014, Roy-Choudhury_2015, Nazir_2016, Iles-Smith_2017, Iles-Smith_2017a, del_Pino_2018, Clear_2020, Svendsen_2023}.
Direct application of the QRT fails to describe such scenarios even qualitatively, missing essential physical features such as phonon sidebands (PSBs)~\cite{McCutcheon_2016, Cosacchi_2021}.
It is often assumed that addressing this requires incorporating non-Markovian effects, either through extensions of the QRT~\cite{Budini_2008, Goan_2011, McCutcheon_2016}, transformations to dressed representations (e.g. polaron)~\cite{Iles-Smith_2017a, Iles-Smith_2017,Bundgaard-Nielsen_2021}, or sophisticated numerical methods~\cite{Strathearn_2018,Jorgensen_2019,Link_2024, Cosacchi_2021, Cosacchi_2018}.

Here we show, to the contrary, that the limitations of the QRT can be overcome using only a weak-coupling Markovian master equation when combined with the sensor approach.
Our key insight is to trace out the structured (vibrational) environment in the \emph{joint system-sensor eigenbasis}, so that the sensors respond directly to both environment-induced transitions and intrinsic system dynamics.
This yields a unified and fully Markovian treatment of frequency-resolved $N$-photon correlations in structured vibrational environments.
Crucially, our approach retains the conceptual simplicity and flexibility of both the sensor and master equation formalisms, while avoiding numerically intensive computations. 

We apply our method to model electron-phonon interactions in realistic solid-state quantum emitters, where it captures key phonon-induced features in the emission spectrum—including PSBs—missed by conventional weak-coupling QRT calcuations~\cite{McCutcheon_2016, Cosacchi_2021}.
Although PSBs in fluorescence are often attributed to non-Markovian effects~\cite{McCutcheon_2016, Cosacchi_2018}, our results show that they can emerge within a weak-coupling Markovian framework.
Most importantly, the computational efficiency of our approach makes the study of phonon influences on frequency-resolved multi-photon correlations tractable, a regime  previously inaccessible.
As an illustration, we compute the emitter two-photon spectrum~\cite{delValle_2012, Gonzalez-Tudela_2013}, which probes cross-correlations between photons at different frequencies.
We uncover a previously unreported phonon-induced feature, showing that signatures of the characteristic correlation structure of the Mollow triplet~\cite{Schrama_1992, Aspect_1980, Dalibard_1983, Arnoldus_1984, Knoll_1986} persist even in photons emitted via the PSB.

\begin{figure}
    \centering
    \includegraphics[width=\linewidth]
    {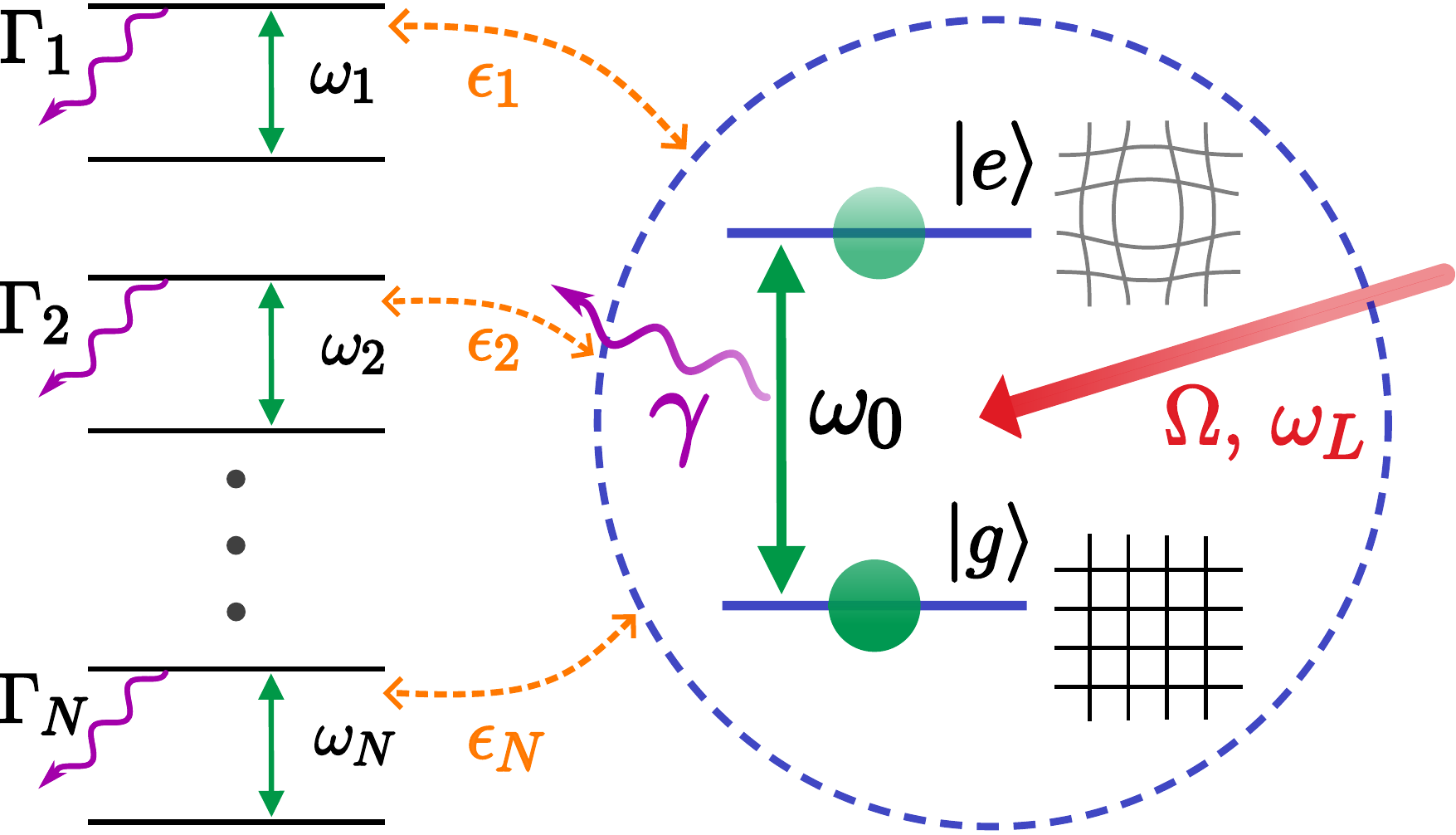}
    \caption{A schematic of the sensor method. $N$~sensors are weakly coupled to a quantum emitter of energy gap $\omega_0$ and decay rate $\gamma$ (encircled), with strength $\epsilon_m$ for sensor $m$. Each sensor is a two-level system with energy gap $\omega_m$ and decay rate $\Gamma_m$. When including vibrations, the quantum emitter is coupled to the host lattice shown to the right. The emitter is externally driven by a laser of frequency $\omega_L$, with Rabi frequency $\Omega$.}
    \label{fig:diagram}
\end{figure}

\textit{The sensor method}---We begin by outlining the sensor method of Ref.~\cite{delValle_2012}, which applies in the absence of any structured vibrational environment.
Suppose a quantum emitter is described by some system Hamiltonian $H_S$ and we are interested in the properties of the radiation it emits.
The time evolution of the emitter's reduced density matrix $\rho_S (t)$ is assumed to follow a master equation of the form $\frac{\partial}{\partial t}\rho_S(t) = \mathcal{L}_0 \rho_S(t)$, with $\mathcal{L}_0 \rho_S = -i[H_S, \rho_S] + \frac{\gamma}{2}\mathcal{D}_{\sigma} (\rho_S)$.
Here $\sigma$ is the lowering operator between the relevant emitter states,  $\mathcal{D}_{\sigma} (\rho)= 2\sigma\rho \sigma^\dagger - \sigma^\dagger \sigma\rho - \rho \sigma^\dagger \sigma$ is a Lindblad dissipator, and $\gamma$ is the emission rate. 

Within the sensor method, frequency-resolved $N$-photon correlation functions of the emitted light are obtained by introducing $N$ \emph{sensors} into the system description (see Fig.~\ref{fig:diagram}).
Sensor $m$ is a two-level system of energy gap $\omega_m$ ($\hbar=1$ throughout), Hamiltonian $H_m = \omega_m\varsigma_m^\dagger\varsigma_m$, lowering operator $\varsigma_m$, and decay rate $\Gamma_m$.
Each sensor is \emph{weakly} coupled to the emitter through 
$H_{S-m} = \epsilon_m(\sigma\varsigma_m^\dagger + \sigma^\dagger \varsigma_m)$, where $\epsilon_m$ is the coupling strength.
The master equation describing the \emph{composite} state of the emitter and  sensors, $\rho(t)$, is then
\begin{equation}
\label{eq:sensor_me}
    \frac{\partial}{\partial t} \rho(t) = \mathcal{L}_0\rho(t) + \sum_{m=1}^N(\mathcal{L}_m \rho(t) - i [H_{S-m}, \rho(t)]),
\end{equation}
with $\mathcal{L}_m \rho = -i[H_m, \rho] + \frac{\Gamma_m}{2}\mathcal{D}_{\varsigma_m} (\rho)$. 
In Ref. \cite{delValle_2012} it was shown that if the sensor couplings are sufficiently weak not to perturb the emitter's dynamics, the sensors act as  detectors with Lorentzian profiles of bandwidth $\Gamma_m$ centred on $\omega_m$.
Frequency-resolved correlation functions of the emitted field then reduce to simple expectation values of sensor operator products. In particular, the steady-state (${\rm ss}$) physical $N$-photon spectrum is given by \cite{delValle_2012}
\begin{equation}
    \label{eq:n_photon_spectrum}
    S^{(N)}_{\Gamma_1, \dots, \Gamma_N}(\omega_1, \dots, \omega_N) = \frac{1}{(2\pi)^N}\frac{\Gamma_1 \cdots \Gamma_N}{\epsilon_1^2 \cdots \epsilon_N^2} \braket{:n_1 \cdots n_N:}_{\rm ss}
\end{equation}
where $n_m = \varsigma^\dagger_m \varsigma_m$ is the number operator of sensor $m$, and $:\,:$ indicates normal  ordering \cite{Vogel_2006}.
As proved in Ref.~\cite{delValle_2012}, this yields results \emph{equivalent} to the QRT once finite frequency resolution is included in the latter.

\textit{Structured environments beyond the QRT}---We now move on to the main development of this work and describe how the sensor approach can be leveraged to incorporate a structured environment {\it beyond} the QRT.
For concreteness, we focus on a solid-state quantum emitter interacting with a vibrational environment formed by its host lattice~\cite{Krummheuer_2002, Nazir_2016}, modelled as a bath of harmonic oscillators.
Each oscillator corresponds to a phonon mode with bosonic annihilation operator $b_k$ and energy $\nu_k$, giving the Hamiltonian $H_E = \sum_k \nu_k b_k^\dagger b_k$.
The emitter-phonon interaction is of the form $H_{S-E} = A \sum_k g_k (b^\dagger_k + b_k)$, where the system operator $A$ couples to mode $k$ with strength $g_k$, characterised by the spectral density $J_{ph}(\nu) = \sum_k \vert g_k\vert^2 \delta(\nu-\nu_k)$, which exhibits non-trivial frequency dependence~\cite{Nazir_2016}.

One could attempt to include this additional environment directly in Eq.~(\ref{eq:sensor_me}) by adding a dissipator.
However, this fails for structured environments~\cite{Mitchison_2018, Maguire_2019, Gribben_2022}, since the proof of the original sensor method establishes its equivalence with a QRT-based approach~\cite{delValle_2012}.
Such an additive treatment would only be valid if the phonon spectral density was flat and the dissipator of Lindblad form, i.e.~under the assumptions of the QRT~\cite{Swain_1981, Budini_2008, Goan_2011, Breuer_2002}.
Key physical features such as PSBs would then be missed.

Instead, we propose an alternative approach: we treat the sensors as part of an extended system and derive a master equation tracing over the phonon environment in the eigenbasis of the \emph{composite} emitter-sensor system.
Unlike the additive treatment, this ensures the sensors respond to energy exchanges between the emitter and phonons.
Moreover, though the \emph{original} sensor method is equivalent to the QRT, the steady-state expectation value in Eq.~(\ref{eq:n_photon_spectrum}) is evaluated without invoking it.
Our framework therefore enables computation of $N$-photon spectra without the limitations of the QRT.

To test this idea and demonstrate the power of our approach, we choose to describe the phonon interactions with a weak-coupling Born-Markov theory~\cite{Nazir_2016, Breuer_2002, Machnikowski_2004, Nazir_2008}.
While this treatment accurately describes emitter dynamics in the weak phonon coupling and low temperature regime~\cite{Nazir_2016, McCutcheon_2010}, it is often assumed to be insufficient for capturing PSBs in the emission spectrum~\cite{McCutcheon_2016, Iles-Smith_2017a}.
Using our framework, we challenge this assumption and show that the failure to reproduce PSBs arises not from the Markovian weak-coupling theory itself, but from the approximations introduced by the QRT.

To derive the weak-coupling master equation we move into the interaction picture with respect to the combined emitter-sensor system Hamiltonian 
$
    H_S^\prime = H_S + \sum_{m=1}^N (H_m + H_{S\text{-}m})
$.
This gives
\begin{equation}
\label{eq:int_picture_transformation_sensor}
    \tilde{H}_{S\text{-}E}(t) = 
    \tilde{A}(t)\sum_k g_k \left(b_k^\dagger e^{i\nu_k t} + b_k e^{-i\nu_k t} \right),
\end{equation}
where, importantly, the transformed operator $\tilde{A}(t) = e^{iH_S^\prime t} A e^{-iH_S^\prime t}$ acts within the Hilbert spaces of both the emitter and sensors.
Following the standard weak-coupling prescription~\cite{Breuer_2002, Nazir_2016,Nazir_2008}, we obtain a master equation of the form (for further details, see \cite{Supplement})
\begin{equation}
\label{eq:extended_sensor_me}
    \frac{\partial}{\partial t} \rho(t) = \mathcal{L}_0\rho(t) + \sum_{m=1}^N(\mathcal{L}_m \rho(t) - i [H_{S-m}, \rho(t)]) + \mathcal{K}(\rho(t)) ,
\end{equation}
where the phonon dissipator $\mathcal{K} (\rho)$ is defined through a rate operator $Z$ as $\mathcal{K} (\rho) = - [ A, Z \rho ] + [ A, \rho Z^{\dagger} ]$.
Using the spectral decomposition of the joint Hamiltonian $H^\prime_S = \sum_\alpha \varepsilon_\alpha \ket{\psi_\alpha}\bra{\psi_\alpha}$, the rate operator is
\begin{equation}
    Z = \sum_{\alpha, \beta} \braket{\psi_\alpha|A|\psi_\beta} \ket{\psi_\alpha}\!\bra{\psi_\beta} \int\limits_0^{\infty} d\tau e^{-i(\varepsilon_\alpha - \varepsilon_\beta)\tau} C(\tau),
\end{equation}
with phonon correlation function $C(\tau) = \int_0^{\infty} d\nu J_{ph}(\nu) \left( \coth{\frac{\beta \nu}{2}} \cos{\nu \tau} - i \sin{\nu \tau} \right)$ at inverse temperature $\beta^{-1} = k_B T$.
We stress again that because $H_S^\prime$ is used rather than $H_S$ in Eq.~(\ref{eq:int_picture_transformation_sensor}), the phonon dissipator $\mathcal{K}$ acts on the composite emitter-sensor Hilbert space.
The $N$-photon spectrum is obtained by setting the left-hand side of Eq.~(\ref{eq:extended_sensor_me}) to zero and solving for the steady state $\rho_{\mathrm{ss}}$ of the emitter-sensor system, which is then used in Eq.~(\ref{eq:n_photon_spectrum}) to compute $S^{(N)}_{\Gamma_1, \dots, \Gamma_N}$.

\begin{figure}
    \centering
    \includegraphics[width=1\linewidth]
    {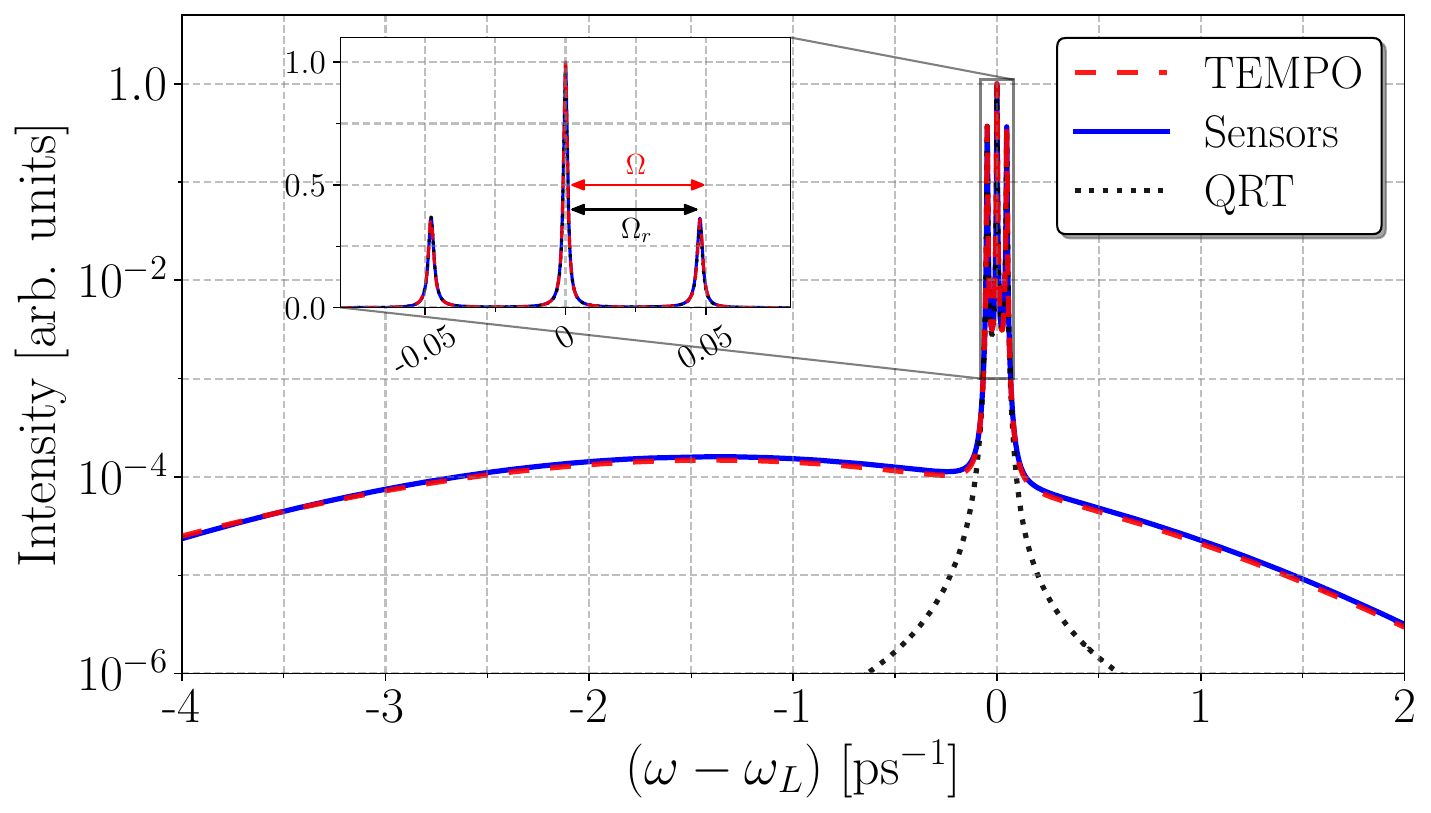}
    \caption{Emission spectrum of a resonantly-driven QD under the influence of a phonon environment calculated using the QRT (dotted, black curves), the sensor-based method introduced in this work (solid, blue curves) and the numerically-exact TEMPO algorithm (dashed, red curves). The inset shows the Mollow triplet in more detail, where $\Omega$ and $\Omega_r$ are the original and phonon-renormalised Rabi frequencies, respectively. Parameters used for this plot are: $\gamma=1/700 \,\textrm{ps}^{-1}$, $\Omega = 0.05 \,\textrm{ps}^{-1}$,  $\alpha=0.027 \,\textrm{ps}^{2}$, $T = 4\,\textrm{K}$ and $\nu_c = 2.2 \,\textrm{ps}^{-1}$. The sensor width is $\Gamma = 10^{-4} \,\textrm{ps}^{-1}$ and the coupling is $\epsilon = 10^{-6} \,\textrm{ps}^{-1}$.}
    \label{fig:sensor_qrt_tempo}
\end{figure}

\textit{Phonon influence on QD emission}---As an example, we apply 
our approach to a semiconductor quantum dot (QD) modelled as a two-level system with ground state $\ket{g}$ and excited state $\ket{e}$, separated by an energy gap $\omega_0$.
The QD is driven resonantly by a continuous-wave laser of frequency $\omega_L$.
In the rotating frame of the laser, and under the rotating-wave approximation, the system Hamiltonian is $H_S = \delta \ket{e}\bra{e} + \frac{\Omega}{2} \sigma_x$, where $\Omega$ is the Rabi frequency, $\delta = \omega_0 - \omega_L$ is the laser detuning, and $\sigma_x = \sigma+\sigma^\dagger$.
Radiative decay with natural linewidth $\gamma$ is included via the Lindblad dissipator $\frac{\gamma}{2} \mathcal{D}_{\sigma}$ in ${\mathcal L}_0$ {\footnote{The radiative decay of the QD can be included additively, since the spectral density of the electromagnetic environment is well approximated as frequency independent and the dissipator is thus of Lindblad type~\cite{PhysRevLett.110.217401,Maguire_2019,Gribben_2022}.}}.
Electron-phonon coupling is through the operator $A = \sigma^\dagger\sigma$, with the phonon environment described by the super-Ohmic spectral density $J_{ph}(\nu) = \alpha\nu^3\exp(-\nu^2/\nu_c^2)$, where $\alpha$ is the coupling strength and $\nu_c$ is the cut-off frequency \cite{Nazir_2016}.

\begin{figure*}
    \centering
    \includegraphics[width=\linewidth]
    {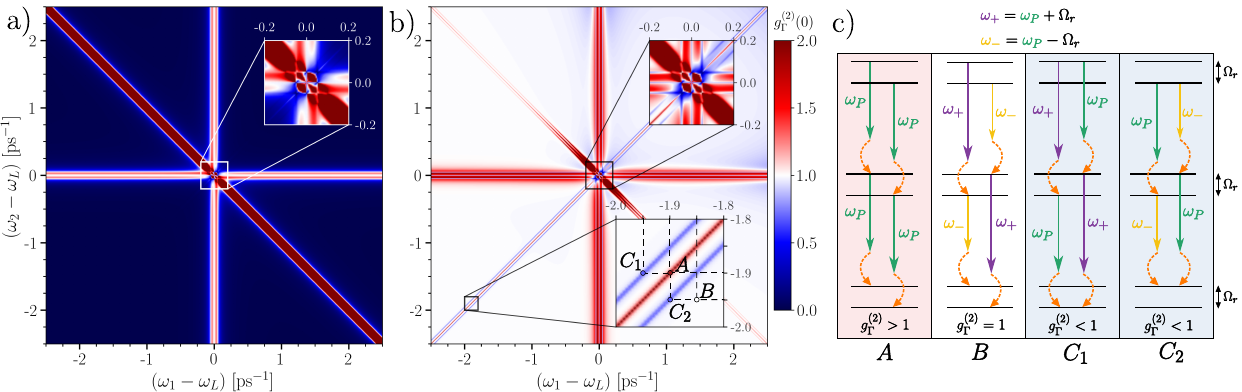}
    \caption{Two-photon spectrum of a resonantly driven QD calculated (a) without and (b) with phonon interactions, where blue corresponds to antibunching, red to bunching, and white to uncorrelated emission. A new feature appears along the diagonal when phonon interactions are included, namely the additional three parallel lines shown in more detail in the lower inset of panel (b). The upper insets in (a) and (b) present magnifications over a smaller range of frequencies close to resonance. Panel (c) gives ladder diagrams in the driven-dot dressed-state basis  showing all the possible photon {\it cascades} involving two phonons of the same energy (orange, dashed), where $\omega_P$ represents an arbitrary photon energy. The letters below the four boxes correspond to the annotated points in the lower inset of panel (b). All parameters are the same as in Fig.~\ref{fig:sensor_qrt_tempo} apart from the sensor width, which is $\Gamma = 2\gamma$ for both sensors. The colorbar scale only extends up to $g^{(2)}_\Gamma(0) = 2$ as this best showcases the phonon-induced triplet structure in the lower inset of Fig.~3~(b). For completeness, the Supplemental Material~\protect{\cite{Supplement}} includes plots with an extended logarithmic colorbar scale.}
    \label{fig:2d_spectrum}
\end{figure*}

We compute the QD physical emission spectrum $S^{(1)}_{\Gamma}(\omega)$ using Eqs.~(\ref{eq:n_photon_spectrum}) and~(\ref{eq:extended_sensor_me}) with $N=1$ sensor.
Fig.~\ref{fig:sensor_qrt_tempo} compares results from our approach (solid blue), the standard QRT (dotted black), and the numerically exact TEMPO algorithm~\cite{Strathearn_2018, Jorgensen_2019, Link_2024} (dashed red).
To obtain the physical spectrum, the QRT and TEMPO predictions are convolved with a Lorentzian of width $\Gamma$~\cite{Eberly_1977}, matching the sensor bandwidth. Importantly, this is only practicable for the single-photon spectrum.
The driving strength is chosen well above saturation ($\Omega = 35\gamma$), fully resolving the Mollow triplet~\cite{Mollow_1969}.

As shown, our method captures the PSB in quantitative agreement with the TEMPO benchmark, while the QRT completely fails. This resolves a long-standing issue: even in the Markovian weak-coupling regime, the PSB can be recovered once the QRT is avoided.
The PSB originates from the Franck–Condon principle~\cite{Besombes_2001, Iles-Smith_2017}, where photon emission or absorption is accompanied by a lattice vibration.
This energy-exchange process is clearly missed by the QRT but captured by our approach.
The inset shows the expected phonon-induced renormalisation of the Mollow sidepeaks~\cite{Ramsay_2010a, Wei_2014}, reproduced correctly by all methods for the parameters considered.

It is worth emphasizing that our approach remains conceptually and computationally simpler than TEMPO.
For a two-level system, the single-photon spectrum is obtained by diagonalizing a $16 \times 16$ matrix for each frequency.
Moreover, it naturally extends to incorporating phonon effects in higher-order photon correlations by simply including additional sensors. 
To our knowledge, such calculations have not been performed previously, as evaluating the required multi-dimensional integrals is prohibitively difficult; neither state-of-the-art numerical methods such as TEMPO nor advanced analytical techniques can currently access these frequency-resolved higher-order correlations in the presence of vibrational environments.

To showcase this capability, we compute the normalized two-photon spectrum $g^{(2)}_{\Gamma_1, \Gamma_2}(\omega_1, \omega_2) = S^{(2)}_{\Gamma_1, \Gamma_2} (\omega_1, \omega_2) / [S^{(1)}_{\Gamma_1}(\omega_1)S^{(1)}_{\Gamma_2}(\omega_2)]$~\cite{Gonzalez-Tudela_2013, delValle_2012} by including two sensors with variable frequencies $\omega_1$ and $\omega_2$.
Experimentally, this quantity can be measured in a coincidence setup using a Hanbury–Brown–Twiss arrangement with  tunable spectral filters ~\cite{Brown_1956, Peiris_2015}.
We choose equal sensor widths $\Gamma_1 = \Gamma_2 = \Gamma = 2\gamma$, so that the Mollow peaks remain well separated ($\Gamma \ll \Omega$) 
while each is fully resolved. Figs.~\ref{fig:2d_spectrum}~(a) and~(b) show the calculated two-photon spectra without and with phonon interactions, respectively.
In both panels red corresponds to bunching ($g_\Gamma^{(2)}>1$), blue to antibunching ($g_\Gamma^{(2)}<1$), and white to uncorrelated emission ($g_\Gamma^{(2)}=1$).
The frequency range extends beyond that typically considered~\cite{Gonzalez-Tudela_2013, Peiris_2015, Carreno_2017}, since phonon-induced features persist at large detunings.
In fact, comparing the two spectra reveals clear signatures of phonon effects across all frequencies shown.
For example, well below resonance, the background exhibits antibunching in the absence of phonon interactions [Fig.~\ref{fig:2d_spectrum}~(a)], 
but becomes mostly uncorrelated with phonon interactions present [Fig.~\ref{fig:2d_spectrum}~(b)].
Without phonons, emission of more than one photon is highly unlikely far from resonance, producing the apparent antibunching.
With phonons, however, there is non-negligible intensity at the same frequencies, with no correlations expected between arbitrary parts of the PSB.

Despite this, when the sensors are tuned near each other but detuned from the emitter, a surprising triplet structure emerges in the presence of phonons, highlighted in the lower inset of Fig.~\ref{fig:2d_spectrum}~(b).
The three peaks are separated by exactly the phonon-renormalized Rabi frequency \cite{Ramsay_2010a, Wei_2014}, $\Omega_r$, matching the sidepeak separations of the Mollow triplet (see Fig.~\ref{fig:sensor_qrt_tempo} inset).
In the absence of phonons, the two-photon correlations of the Mollow triplet at zero delay are well understood~\cite{Aspect_1980, Dalibard_1983, Schrama_1992, Nienhuis_1993, Gonzalez-Tudela_2013, Carreno_2017}.
Photon pairs from the central peak are uncorrelated, as are pairs from the opposite sidepeaks (separated by twice the Rabi frequency).
In contrast, photon pairs from the central peak and one sidepeak (separated by the Rabi frequency), or two from the same sidepeak, are antibunched.

Remarkably, our results show that signatures of this correlation structure persist within the phonon sideband, indicating that while phonons open additional emission pathways, they do not fundamentally alter the underlying second-order coherence of the Mollow triplet.
Specifically, within the PSB, photon pairs differing in frequency by $\Omega_r$ appear antibunched, giving rise to blue diagonal lines in Fig.~\ref{fig:2d_spectrum}~(b).
Pairs separated by $2\Omega_r$
show no correlation (invisible against the white background), while identical-frequency pairs exhibit strong bunching (red diagonal line) as a consequence of frequency filtering causing the \emph{indistinguishability bunching} effect~\cite{Gonzalez-Tudela_2013}, where $g^{(2)}_{\Gamma,\Gamma}(\omega,\omega) \to 2$ for $\Gamma \to 0$, masking any other underlying correlations.
Fig.~\ref{fig:2d_spectrum}~(c) maps out the corresponding two-photon cascades in the dressed-state basis~\cite{Cohen-Tannoudji_1977}, where each photon emission is accompanied by a fixed-energy phonon emission.
Box A depicts identical-frequency pairs, bunched due to indistinguishability.
Pairs differing by $\Omega_r$ (boxes $C_1$ and $C_2$) can occur in two time orderings that interfere destructively, producing antibunching.
This interference stems from the finite detector time resolution ($\sim 1/\Gamma$), which introduces uncertainty in emission ordering~\cite{Schrama_1991, Schrama_1992, Nienhuis_1993, Gonzalez-Tudela_2013}.
Pairs differing by $2\Omega_r$ (box $B$) appear uncorrelated, consistent with the zero-delay behaviour of opposite Mollow sidepeaks without phonons~\cite{Schrama_1992, Nienhuis_1993, Gonzalez-Tudela_2013}.

\textit{Conclusions}—We have introduced a general framework for calculating frequency-resolved $N$-photon spectra in the presence of structured vibrational environments, overcoming the limitations of the quantum regression theorem.
By embedding electron–phonon interactions directly into a weak-coupling master equation for a combined emitter–sensor system, we capture phonon-assisted transitions inaccessible to the QRT~\cite{McCutcheon_2016, Cosacchi_2021}.
Applied to semiconductor quantum dots, our approach reproduces numerically exact results with high accuracy, including the phonon sideband.
This shows that the failure of previous QRT-based treatments to account for sidebands stems from the QRT itself, not from the weak-coupling master equation.
Leveraging the flexibility of the sensor formalism, we also compute higher-order correlations and, for the first time, investigate the phonon influence on the two-photon spectrum.
We uncover a distinct diagonal triplet structure arising from phonon-assisted emission and show that hallmark correlations of the Mollow triplet persist even within the phonon sideband.

Although we have focused on a weak-coupling master equation, our formalism is agnostic to this choice.
More advanced treatments, such as the polaron transformation~\cite{McCutcheon_2010, Nazir_2016}, could be incorporated to explore stronger coupling regimes.
Extending this approach to temporal correlations presents another promising direction.
We anticipate that our predictions, particularly the phonon-assisted two-photon features, will motivate experimental efforts to observe and exploit these effects.

\bibliography{references_sensors_with_phonons}

\pagebreak
\widetext
\begin{center}
\textbf{\large Supplemental Material: ``A Markovian approach to $N$-photon correlations beyond the quantum regression theorem"}
\end{center}

\setcounter{equation}{0}
\setcounter{figure}{0}
\setcounter{table}{0}
\setcounter{page}{1}
\makeatletter
\renewcommand{\theequation}{S\arabic{equation}}
\renewcommand{\thefigure}{S\arabic{figure}}

\begin{center}
Here we present details on the derivation of the phonon dissipator, some additional analysis of two-photon spectra, and an outline of the TEMPO calculations.
\end{center}

\section{Weak-coupling master equation including sensors}

We start with the derivation of the weak-coupling master equation, tracing over the vibrational environment in the joint emitter-sensor eigenbasis. We consider a Hamiltonian of the system and the vibrational environment (without sensors) of the form
\begin{equation}
    \label{eq:base_hamiltonian}
    H_0 = H_S + \underbrace{A \sum_k g_k (b^\dagger_k + b_k)}_{H_{S-E}} + \underbrace{\sum_k \nu_k b^\dagger_k b_k}_{H_E} ,
\end{equation}
where $H_S$ and $A$ are arbitrary operators acting on the system, and $b_k$ is the bosonic annihilation operator for phonon mode $k$, with frequency $\nu_k$ and system coupling strength $g_k$.

As described in the main text, in order to include the sensors in the description we define the joint emitter-sensor Hamiltonian $H_S^\prime$, which is given in the frame rotating with the laser frequency $\omega_L$ by
\begin{equation}
    H^\prime_\textrm{S} = H_S + \sum_m (\omega_m - \omega_L) \varsigma_m^\dagger \varsigma_m + \epsilon_m (\sigma^\dagger \varsigma_m + \sigma \varsigma_m^\dagger).
\end{equation}
The total Hamiltonian of the emitter, sensors, and the phonon environment is thus $H= H^\prime_S + H_{S-E} + H_E$.

The time evolution of the full system-environment state $\chi(t)$ is governed by the Liouville-von Neumann equation $\frac{\partial}{\partial t}\chi(t) = -i [H, \chi(t)]$. As outlined in the main text, we now move to the interaction picture with respect to $H^\prime_S + H_E$, noting that this includes the sensors in the transformation. The Liouville-von Neumann equation in the interaction picture is now
\begin{equation}
    \label{eq:liouville-vonNeumann}
    \frac{\partial}{\partial t}\tilde\chi(t) = -i [\tilde{H}_{S-E}(t), \tilde\chi(t)],
\end{equation}
where the interaction Hamiltonian in the interaction picture is given by Eq.~(3) in the main text, that is
\begin{equation}
    \label{eq:intpic_interaction_Hamiltonian}
    \tilde{H}_{S\text{-}E}(t) = \left( e^{iH_S^\prime t} A e^{-iH_S^\prime t} \right) \sum_k g_k \left(b_k^\dagger e^{i\nu_k t} + b_k e^{-i\nu_k t} \right).
\end{equation}
Crucially, since $H^\prime_S$ with sensors included is used to define the interaction picture, rather than the bare $H_S$, the transformed system operator $\tilde{A}(t) = e^{iH_S^\prime t} A e^{-iH_S^\prime t}$ acts within the Hilbert spaces of both the emitter and sensors.

The derivation of the weak-coupling master equation then follows the usual prescription as found for example in Sec.~III of~\cite{Nazir_2016} or in~\cite{Breuer_2002}. Thus, we can simply write down the standard Born-Markov master equation \cite{Nazir_2016, Breuer_2002} for the composite emitter-sensor state $\rho(t)$ in the interaction picture as
\begin{equation}
    \frac{\partial}{\partial t} \tilde\rho (t) = - \int_0^{\infty} d\tau \left( \left[ \tilde{A}(t),  \tilde{A}(t-\tau) \tilde\rho (t) \right] C (\tau) -  \left[ \tilde{A}(t),  \tilde\rho (t) \tilde{A}(t-\tau) \right] C (- \tau) \right) ,
\end{equation}
where the phonon environment correlation function at inverse temperature $\beta = 1/k_B T$ is given by
\begin{equation}
    C (\tau) = \int_0^{\infty} d\nu J_{ph}(\nu) \left( \coth{\frac{\beta \nu}{2}} \cos{\nu \tau} - i \sin{\nu \tau} \right),
\end{equation}
with $J_{ph}(\nu)$ being the phonon spectral density. Transforming back to the Schrödinger picture, we obtain
\begin{equation}
    \frac{\partial}{\partial t} \rho (t) = -i[H^\prime_S, \rho(t)] - \int_0^{\infty} d\tau \left( \left[ A,  A(-\tau) \rho (t) \right] C (\tau) -  \left[ A,  \rho (t) A(-\tau) \right] C (- \tau) \right) .
\end{equation}
We may further define the rate operator
\begin{equation}
\label{eq:z_rate_operator}
    Z \equiv \int_0^\infty d\tau C(\tau) A(-\tau)
\end{equation}
and, assuming that $A$ is Hermitian, write the master equation as
\begin{equation}
\label{eq:master_eq_with_phonon_dissipator}
    \frac{\partial}{\partial t} \rho (t) = - i [ H^\prime_S, \rho (t) ] \underbrace{- [ A, Z \rho (t) ] + [ A, \rho (t) Z^{\dagger} ]}_{\mathcal{K}(\rho(t))} ,
\end{equation}
where the last two terms form the phonon dissipator $\mathcal{K}(\rho(t))$ acting on the joint system-sensor Hilbert space.

The phonon dissipator can be expressed using the eigendecomposition
\begin{equation}
\label{eq:eigendecomposition}
    H^\prime_S = \sum_\alpha \varepsilon_\alpha \ket{\psi_\alpha}\bra{\psi_\alpha} ,
\end{equation}
where $\{\varepsilon_\alpha\}$ are the eigenvalues and $\{\ket{\psi_\alpha}\}$ the eigenvectors of $H^\prime_S$.
The system interaction operator in the interaction picture can then be given as
\begin{equation}
\label{eq:int_pic_operator_eigenvalues}
    A(t) = \sum_{\alpha,\beta} A_{\alpha\beta} \ket{\psi_\alpha}\bra{\psi_\beta} e^{i\lambda_{\alpha\beta}t},
\end{equation}
where we have defined $\lambda_{\alpha\beta} = \varepsilon_\alpha - \varepsilon_\beta$ and $A_{\alpha\beta} = \braket{\psi_\alpha|A|\psi_\beta}$. Plugging Eq.~(\ref{eq:int_pic_operator_eigenvalues}) into Eq.~(\ref{eq:z_rate_operator}) we obtain a new expression for the phonon rate operator, equivalent to Eq.~(5) in the main text,
\begin{equation}
\label{eq:z_rate_operator_eigen}
    Z = \sum_{\alpha, \beta} A_{\alpha\beta} \ket{\psi_\alpha}\bra{\psi_\beta} \mathcal{F}(-\lambda_{\alpha\beta})
\end{equation}
where $\mathcal{F}(\lambda)$ is the one-sided Fourier transform of the correlation function
\begin{equation}
\label{eq:fourier_transform_phonon_correlation}
    \mathcal{F} (\lambda) = \int_0^{\infty} d\tau e^{i\lambda\tau} C(\tau) .
\end{equation}
Eqs.~(\ref{eq:z_rate_operator_eigen}) and (\ref{eq:fourier_transform_phonon_correlation}) can be inserted into Eq.~(\ref{eq:master_eq_with_phonon_dissipator}) together with the eigendecomposition of $H^\prime_S$ to obtain the final form of the phonon dissipator for the joint system-sensor Hilbert space.

Finally, we add to the master equation the sensor decay (with rate $\Gamma_m$) and the radiative decay of the emitter (with rate $\gamma$), such that the full master equation is
\begin{equation}
\label{eq:sensor_phonon_me}
    \frac{\partial}{\partial t} \rho (t) = -i [H^\prime_S, \rho(t)] + \mathcal{K} (\rho(t)) + \frac{\gamma}{2}\mathcal{D}_\sigma(\rho(t)) + \sum_m\frac{\Gamma_m}{2} \mathcal{D}_{\varsigma_m} (\rho(t)).
\end{equation}
The radiative decay can simply be added to the obtained master equation when the spectral density of the electromagnetic environment can be considered as frequency-independent over energy scales of interest, with the result that the dissipator is of Lindblad form with a constant rate. The assumptions of the QRT are then satisfied and the additive treatment is valid, as explained in the main text. 

\subsection{Note on the polaron shift}

When deriving the weak-coupling master equation for the phonon environment (without sensors), we obtain a \textit{polaron shift} term in the master equation, given by
\begin{equation}
    -i \delta_P [\sigma^\dagger\sigma, \rho_S(t)],
\end{equation}
where $\rho_S(t)$ is the reduced state of the system (the quantum dot). The polaron shift $\delta_P$ is defined by
\begin{equation}
    \delta_P = \mathrm{Im} \left\{  \int_0^\infty C(\tau) \,\mathrm{d}\tau \right\} = - \int_0^\infty \frac{J_{ph}(\nu)}{\nu} \,\mathrm{d}\nu .
\end{equation}
For the QD phonon spectral density, $J_{ph}(\nu)=\alpha\nu^3\exp(-\nu^2/\nu_c^2)$, the polaron shift is $\delta_P = - \alpha \sqrt{\pi} \frac{\nu_c^3}{4}$.

We can thus calculate the value of the polaron shift and compare it to the energy scales of the problem. If we consider resonance fluorescence, then the detuning $\delta = \omega_0 - \omega_L = 0$ and the energy scale of the system Hamiltonian is given by the Rabi frequency $\Omega$. For the QD and phonon parameters used in the main text, that is $\Omega = 0.05 \,\textrm{ps}^{-1}$, $\alpha = 0.027 \,\textrm{ps}^{2}$ and $\nu_c = 2.2 \,\textrm{ps}^{-1}$, we calculate the polaron shift to be $\delta_P \simeq -0.127 \,\textrm{ps}^{-1}$, which is comparable to the Rabi frequency. As such, it would be incorrect to treat it as part of the perturbation, since it is of the same energy scale as $H_S$.

A simple solution to this problem is to \textit{redefine} what constitutes the system and interaction Hamiltonian in such a way that the polaron shift is removed from $H_{S-E}$ and included instead in $H_S$. Thus, we now have
\begin{equation}
    \label{eq:redefined_hamiltonian}
    H_0 = \underbrace{(\omega_0^\prime-\omega_L)\sigma^\dagger\sigma + \frac{\Omega}{2}\sigma_x}_{H_S} + \underbrace{\sigma^\dagger\sigma \sum_k g_k (b_k^\dagger + b_k) - \delta_P\sigma^\dagger\sigma}_{H_{S-E}} + \underbrace{\sum_k \nu_k b_k^\dagger b_k}_{H_E} ,
\end{equation}
where we have also incorporated the polaron shift into the definition of the QD energy gap $\omega_0^\prime = \omega_0 + \delta_P$. Thus, resonance fluorescence is obtained when the laser is tuned to the \emph{observed} resonance peak ($\omega_L = \omega_0^\prime$), in agreement with typical experimental procedures. In numerical calculations this means that we can set $\omega_0^\prime - \omega_L = 0$ to drive resonantly with the polaron-shifted transition frequency~\cite{Nazir_2016}.

The Born-Markov master equation for this modified Hamiltonian in Eq.~(\ref{eq:redefined_hamiltonian}) is identical to the one obtained for unmodified $H_0$ in Eq.~(\ref{eq:base_hamiltonian}) aside from an additional term that exactly compensates for the polaron shift. Thus, this simple redefinition of the QD energy gap suffices to remove the polaron shift from the master equation.

\section{Full-scale two-photon spectra}

\begin{figure}[h]
    \centering
    \includegraphics[width=0.65\linewidth]{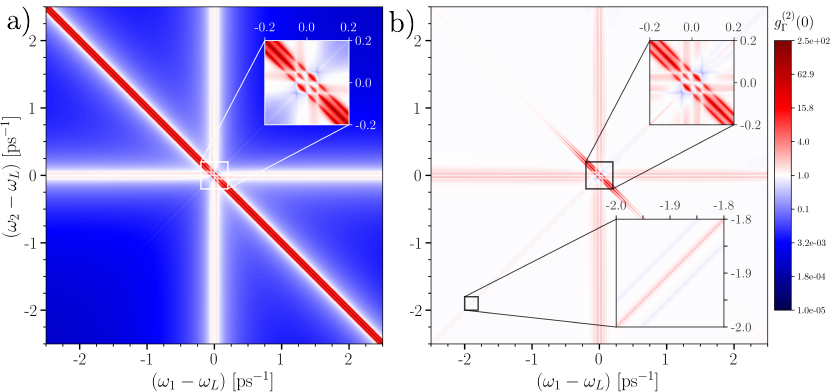}
    \caption{Two-photon spectra of a resonantly-driven QD calculated (a) without and (b) with phonon interactions, as in Fig.~3 of the main text. The colorbar scale is logarithmic above and below $g_{\Gamma}^{(2)}(0) = 1$ and covers the whole range of values present in both plots. The phonon-induced triplet feature is still present, with strong bunching features due to leapfrog processes in the upper insets of both plots now better resolved.}
    \label{fig:TPS_full_scale}
\end{figure}

The colorbar used in Fig.~3~(a) and~(b) only covers the range of values $0 < g^{(2)}_{\Gamma_1, \Gamma_2}(\omega_1, \omega_2) < 2$, since phonon-induced features in the two-photon spectrum are best showcased with this choice. All the values in the lower inset of Fig.~3~(b) are indeed contained within this range.

However, this choice means that values of $g^{(2)}_{\Gamma_1, \Gamma_2}(\omega_1, \omega_2) > 2$ appear as having the value of $g^{(2)}_{\Gamma_1, \Gamma_2}(\omega_1, \omega_2) = 2$. This masks some of the features close to resonance, such as strong bunching for the \emph{leapfrog} processes~\cite{Gonzalez-Tudela_2013}.

In Fig.~\ref{fig:TPS_full_scale} we thus show the same two-photon spectra data as in the main text, but with a different colorbar scale that covers the whole range of $g^{(2)}_{\Gamma_1, \Gamma_2}(\omega_1, \omega_2)$ values present in the data. The main features discussed in the paper are not affected while the upper insets now show the expected strong bunching associated with  leapfrog processes. 

\section{Numerically-exact calculations with TEMPO}

The numerically exact results we use to demonstrate the accuracy of the spectra obtained with our method are produced using the uniform time-evolving matrix product operator (uniTEMPO) method~\cite{Link_2024}.
To outline the approach, we obtain a translationally invariant representation of the influence functional for the system evolution, which fully encodes the non-Markovian process generated by the system-environment coupling and system interventions. The corresponding process tensor~\cite{Jorgensen_2019} can then be used to compute the correlation functions of interest, including steady-state calculations.

The algorithm developed by Link \emph{et al.}~\cite{Link_2024} takes as it's input the bath correlation function $C(\tau)$, as defined in the main manuscript.
From this, the influence functional is propagated using an infinite time-evolving block decimation (iTEBD) algorithm, in which each iTEBD gate is constructed from the memory kernel of the environment.  
The resulting influence functional $f_{\delta t}$ is a uniform Matrix Product State (uMPS), which has bond dimension that depends strongly on $\alpha$, $T$ and the simulation time step $\delta t$, which must be chosen to be sufficiently small to satisfy the Trotter decomposition employed. The left and right boundary vectors $v_l$ and $v_r$ are then calculated from each uMPS according to the technique detailed in Ref.~\cite{Link_2024}.

As in this work the system Hamiltonian is time-independent (in the rotating frame), taking any influence functional uMPS we may obtain a uniform matrix product operator representation of the process tensor. To do this, we first calculate the system propagator: $\mathcal{V}^{1/2} = e^{\mathcal{L}_0\delta t/2},$ where $\mathcal{L}_0$ is the Liouville superoperator as defined in the main text. This is then combined with $f_{\delta t}$ according to the trotter decomposition $(\Upsilon_{\delta t})_{a,\beta,b,\psi} = \sum_\varepsilon (f)_{a,\varepsilon,b}(\mathcal{V}^{1/2}_{j,j-1})_{\varepsilon,\beta}(\mathcal{V}^{1/2}_{j-1,j-2})_{\psi,\varepsilon}$, where $a$, $b$, $\beta$, $\psi$ are free indices, to obtain a process tensor site $\Upsilon_{\delta t}$, the system indices are given by Greek letters and environment by Latin.  Functionally, the process tensor is a multilinear map that propagates a system state, taking into account the conditioning of the environment from the past. 

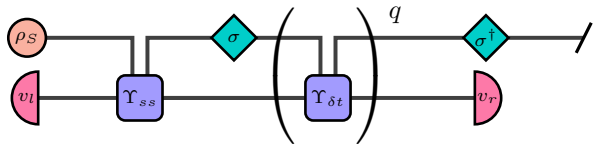
\begin{figure}[ht]
    \centering
    \begin{tikzpicture}
        \Vertex[x=0.5, y=-5.2, size=0.5, label=$\rho_S$,style={fill=tensorred}]{rho0}
        \Vertex[shape=semicircle, style={shape border rotate=90,fill=tensorpink}, x=0.5, y=-6, label=$v_l$, size=.35]{vl}
        \Vertex[style={rounded vertex,fill=tensorpurp}, x=2, y=-6, shape=rectangle, label=$\Upsilon_{ss}$]{upt1}
        \Edge(vl)(upt1)
        \Vertex[style={rounded vertex,fill=tensorpurp}, x=4.5, y=-6, shape=rectangle, label=$\Upsilon_{\delta t}$]{upt2}
        \Edge(upt1)(upt2)
        \Edge[path={{1.9, -5.2}, {1.9, -6}}](rho0)(upt1)
        \Vertex[shape=diamond,style={fill=tensorteal}, x=3.25, y=-5.2, label=$\sigma$]{sigma2}
        \Edge[path={{2.1,-6},{2.1,-5.2}}](upt1)(sigma2)
        \Edge[path={{4.4,-5.2},{4.4,-6}}](sigma2)(upt2)
        \Vertex[shape=diamond,style={fill=tensorteal}, x=6.6, y=-5.2, label=$\sigma^\dagger $]{sigma_s}
        \Vertex[Pseudo, x=8, y=-5.2, size=0]{tr}
        \Edge[path={{4.6,-6},{4.6,-5.2}}](upt2)(sigma_s)
        \node at (4.56, -5.7) {\Large $\left(\phantom{\rule{0.8cm}{1cm}}\right)^{q}$};
        \Vertex[shape=semicircle, style={shape border rotate=-180,fill=tensorpink}, x=6.6, y=-6, label=$\,v_r$, size=.35]{vr}
        \Edge(upt1)(vr)
        \Edge(sigma_s)(tr)
        \draw [-,very thick] (8.0,-5.0) -- (7.8,-5.4);
    \end{tikzpicture}
    \caption{Tensor network representation of the calculation of the correlation function with uniTEMPO. The system and environment is first propagated to the joint steady-state before the first operator $\sigma$ is applied. This is followed by further time evolution. This is denoted with the site $\Upsilon_{\delta t}$ being applied $q$ times (indicated by the brackets and power). Lastly, the second operator and system trace $\sigma^\dagger$ are applied. The boundary vectors $v_l$ and $v_r$ may be viewed as the bath preparation and trace respectively.}
    \label{fig:unitempo_cfn}
\end{figure}

Due to the time-translation invariance and time-independence of the Liouvillian, steady-state calculations can be obtained directly from $\Upsilon_{\delta t}$. With this, the steady state in the correlation function defined in the main text may be computed in a numerically exact fashion. For this, we calculate the steady-state process tensor site $\Upsilon_{ss}$ from the leading eigenvector of $\Upsilon_{\delta t}$ which has the same dimensions as $\Upsilon_{\delta t}$. We can then calculate the desired correlation functions following Fig.~(\ref{fig:unitempo_cfn}). The initial state $\rho_S=\ket e\bra e$ and left boundary vector $v_l$ form the initial conditions. The steady state propagator $\Upsilon_{ss}$ first evolves the combined system and environment to the steady-state, governed by $\mathcal{L}_0$, without a trace on the environment. The first operator $\sigma$ is applied to the system, followed by further time evolution of $q$ time steps giving $\tau=q\delta t$. Finally the second operator $\sigma^\dagger$ is applied to the system before the trace of the system is taken (indicated by the slash) and the right boundary vector is applied. This concept can be generalised to \emph{unfiltered} higher order correlations but with the associated increased computational effort related to the number of time variables that are varied.

To ensure numerical convergence of the computations, we refine the SVD tolerance parameter until any resulting changes have become negligible. We found that a tolerance of $10^{-12}$ with a step size of $\delta t=~0.5$~ps and a memory cutoff of $\tau_c = 15$~ps is sufficient to ensure convergence for all values of $\alpha$ and $T$ required. For the calculation of correlation functions, the number of time steps taken, parameterised by the integer $q$, is chosen to be sufficiently large such that the spectra are converged.

\end{document}